\documentclass[aps,pre,twocolumn,showpacs,floatfix,superscriptaddress]{revtex4}
\usepackage{graphicx}
\usepackage{graphics}
\usepackage[usenames]{color}

\newcommand{\be}{\begin{equation}}
\newcommand{\ee}{\end{equation}}
\newcommand{\bea}{\begin{eqnarray}}
\newcommand{\eea}{\end{eqnarray}}

\begin{document}

\title{Shock Propagation in Granular Flow Subjected to an External 
Impact}
\author{Sudhir N. Pathak}
\email{sudhirnp@imsc.res.in}
\affiliation{Institute of Mathematical Sciences, CIT Campus, Taramani, 
Chennai-600113, India}
\author{Zahera Jabeen}
\email{zjabeen@umich.edu}
\affiliation{ Department of Physics, University of Michigan, Ann Arbor, MI 48109-1040, USA}
\author{Purusattam Ray}
\email{ray@imsc.res.in}
\affiliation{Institute of Mathematical Sciences, CIT Campus, Taramani, 
Chennai-600113, India}
\author{R. Rajesh}
\email{rrajesh@imsc.res.in}
\affiliation{Institute of Mathematical Sciences, CIT Campus, Taramani, 
Chennai-600113, India}
  
\date{\today}

\begin{abstract}
We analyze a recent experiment [Phys. Rev. Lett., {\bf103}, 224501 
(2009)] in which the shock, created by the impact of a steel ball on a 
flowing monolayer of glass beads, is quantitatively studied. We argue 
that radial momentum is conserved in the process, and hence show that in two dimensions the shock radius increases in time $t$ as a power law $t^{1/3}$. This is 
confirmed in event driven simulations of an inelastic hard sphere 
system. The experimental data are compared with the theoretical 
prediction, and is shown to compare well at intermediate times. At late 
times, the experimental data exhibit a crossover to a different scaling 
behavior. We attribute this to the problem becoming effectively three 
dimensional due to accumulation of particles at the shock front, and 
propose a simple hard sphere model which incorporates this effect. Simulations of 
this model capture the crossover seen in the experimental data. 
\end{abstract}

\maketitle

\section{\label{introduction} Introduction}

Driven granular gases can produce complex and intricate spatial patterns 
\cite{jaeger1996,kadanoff1999,kudrolli2004,aranson2006}. Of particular 
interest is pattern formation following a localized perturbation, the 
subject matter of many recent experiments. Examples include crater 
formation by wind jets in the context of lunar cratering 
\cite{metzger}, viscous fingering in grains confined in a 
Hele-Shaw cell when displaced by gas or liquid 
\cite{jaeger2008,sandnes2007,pinto2007}, shock propagation in flowing 
glass beads following a sudden impact \cite{boudet2009}, signal 
propagation in dilute granular gas \cite{losert} as well as in dense 
static granular material (see \cite{luding} and references within), and 
avalanches in sand piles~\cite{daerr}.

In this paper, we focus on an experiment \cite{boudet2009} (henceforth 
referred to as BCK) on a dilute monolayer of glass beads flowing on an 
inclined glass plane.  In the experiment, a steel ball, much larger in 
size than an individual glass bead, is dropped from a height onto the 
flowing beads. The impact generates a circular region, devoid of glass 
beads, whose radius increases with time. This radius was measured using 
high speed cameras. A theoretical model was proposed, and analyzed to 
derive an equation obeyed by the radius, whose solution predicts a 
logarithmic growth at large times. The numerical solution of the 
equation was shown to match with the experimental data 
\cite{boudet2009}.

In an independent study, we had studied the effect of exciting a single 
particle in a system of stationary hard inelastic particles using event 
driven molecular dynamics simulations, and scaling arguments 
\cite{zahera}. By identifying radial momentum as a conserved quantity, 
and using scaling arguments, the radius of disturbance was predicted to 
increase with time as a power law $t^{1/3}$ in two dimensions. This result was shown to be 
in very good agreement with data from numerical simulations of the 
model.

The inelastic hard sphere model closely resembles the experimental 
system in BCK, in the limit when the impact is very intense. In this 
paper, we propose the power law $t^{1/3}$ as an alternate description of 
the radius of disturbance in the BCK experiment.  By reexamining the 
data in BCK, we show that, there are temporal regimes in which the power law growth is a good description. At late times, the experimental data deviate 
from the $t^{1/3}$ behaviour.  This, we argue is due to the experimental 
system becoming effectively three dimensional, and propose a simple 
model incorporating this effect. Our numerical data, obtained from 
simulations of this model, show clearly the crossover and captures the 
long time behaviour. Since these results are in contradiction to those 
presented in BCK, we further analyze the model proposed in BCK, 
and point out some shortcomings. In particular, we show numerically that
the main assumption of BCK is not
correct. Though the experimental data are not able to distinguish 
between the two theories because the time
scales are not large enough, the simulation data clearly bring out the 
deficiencies of the BCK theory at large times.

The experiment in BCK is the inelastic version of the classic Taylor-von 
Neumann-Sedov problem of shock propagation following a localized intense 
explosion \cite{sedov}. In the latter case, the particles remain 
homogeneously distributed and the exponents characterizing the power law 
growth of the radius of the disturbance follows from energy conservation 
and simple dimensional analysis \cite{taylor}, while the scaling 
functions can be calculated exactly following a more detailed analysis 
\cite{sedov,neumann}. Theoretical, numerical and experimental studies of 
this problem are summarized in Refs.~\cite{zeldovich,ostriker}. 
Simulations in a hard sphere model with elastic collisions reproduce the 
results based on scaling arguments \cite{antal2008}.

The BCK experiment is also a special case of a freely cooling gas (in a reference frame moving with mean velocity of particles) 
wherein, after the initial input of energy, the system is isolated and 
allowed to freely evolve without any external driving.
A key feature of the freely cooling granular gas is the clustering due to inelastic collisions. The freely cooling gas is well understood 
in one dimension and progressively less understood as the dimension 
increases 
\cite{haff,carnevale,britoernst,mcnamara_pre,trizac1,trizac2,bennaim1999,nie2002,luding2,goldhirsch1993,shinde2007,shinde2009, 
supravat2011,schen2000,frachebourg,trizac4}. Such systems are 
challenging experimentally because inelasticity is overwhelmed by 
friction and boundary effects. Friction can be eliminated in experiments 
on particles under levitation \cite{maa} or in microgravity 
\cite{haffexperi1,Grasselli2009}, but are expensive to perform and are 
limited by small number of particles and short times. In BCK, friction 
is balanced by gravity, and at high enough impact energies, in the center of mass frame, mimics a 
stationary collection of inelastic particles without friction. The 
boundary effects are eliminated as long as the shock does not reach the 
edges of the container. Thus, it is an experiment where clustering due 
to inelastic collisions can be studied easily.

The rest of the paper is organized as follows. In Sec.~\ref{bckmodel}, 
we describe the theoretical model in BCK, and review the arguments 
that lead to the equation obeyed by the radius of the disturbance. This 
equation is further analyzed to derive the asymptotic long 
time behaviour. The shortcomings of this model are pointed out. We then, in 
Sec.~\ref{model} define a hard core inelastic gas model on which our 
computer simulations are performed. In Sec.~\ref{bckmodel_numerics}, we demonstrate that our  model reproduces 
the basic features of the experiment in BCK. The assumptions of the 
analysis in BCK is tested within this model and counter evidence is 
presented. In Sec.~\ref{comparison}, we compare the experimental results 
in Ref.~\cite{boudet2009} with the power law growth rules obtained in 
Ref.~\cite{zahera}.  The data at intermediate times are well described 
by these power laws. However, there is a crossover to a different 
behaviour at large times. In Sec.~\ref{modifications}, we examine whether 
this large time behaviour can be explained in terms of velocity 
fluctuations of the particles or by making the rim three dimensional. We 
argue that it is plausible that the three dimensional rim is responsible 
for deviation from power law growth and verify this by simulation. The 
results are summarized in Sec.~\ref{conclusion}.

\section{\label{bckmodel}  BCK model and Analysis}

We first review the model studied in BCK to explain the experimental 
data. The model is based on the experimental observation, that after the 
impact with the steel ball, the displaced glass beads form a growing 
circular ring, devoid of beads. BCK considered an idealized model where 
all the particles contained in a disk of radius $R(t)$ at time $t$ 
accumulate at the rim (boundary of ring). The remaining particles that are outside
the disk are assumed to be stationary. This 
mimics the experimental system when one transforms to the center of mass 
coordinates, and in the limit of large impact energy, when the 
fluctuations of the particle velocities about the mean flow may be 
ignored. Each particle at the rim is assumed to move radially outwards 
with a speed $V(t)$. As the ring moves outwards, more particles are 
absorbed into the ring. We reproduce the calculation in BCK, but 
generalized to $d$-dimensions. The total kinetic energy $E(t)$ is
\be
E(t) = \frac{1}{2} \rho_0 \Omega_d R(t)^d V(t)^2,
\label{eq:energy}
\ee
where $\rho_0$ is the initial mass density, and $\Omega_d$ is the volume 
of a unit sphere in $d$-dimensions, such that $\rho_0 \Omega_d R(t)^d$ 
is the total mass of displaced particles. The speed $V(t)$ is
\be
V(t) = \frac{d R(t)}{dt}.
\label{eq:speed}
\ee

One more relation between $E(t)$ and $R(t)$ is required for the 
solution. If the particles were elastic, then total energy is conserved, 
$E(t) \sim t^0$, and one obtains $R(t) \propto t^{2/(d+2)}$; in 
particular, $R(t) \propto \sqrt{t}$ in $d=2$ \cite{taylor}. However, when 
particles are inelastic, there is no such conservation law, and energy 
decreases with time. BCK proceed by the following argument. If $r$ is 
the coefficient of restitution, then the loss of energy when a particle 
in the rim collides with a stationary particle outside is 
$\frac{1}{2}(1-r^2) V(t)^2$. Thus, when the ring moves out by a distance 
$dR$, then the change in energy $dE$ is given by
\be
d E = -\frac{1}{2} \Omega  R(t)^d \rho_0 V(t)^2 (1-r^2) N(t) dR,
\label{energy_evol}
\ee
where $N(t)$ is the number of collisions per particle per unit length, 
or equivalently, $N(t) dR$ is the number of collisions for each particle 
in the rim as it travels a distance $dR$. BCK makes the strong 
assumption that $N(t)$ is independent of the radius, and hence time $t$ 
that is,
\be
N(t) = \rm{constant}.
\label{eq:assumption_bck}
\ee
Eliminating $R(t)$ and $V(t)$ in Eq.~(\ref{energy_evol}) using 
Eq.~(\ref{eq:energy}), one obtains
\be
E(t)= E_0 \exp\left[ -N (1-r^2) R(t) \right],
\label{eq:energy_soln}
\ee
where $E_0$ is the energy of impact at $t=0$. It is now straightforward 
to obtain the equation satisfied by the radius $R(t)$:
\be
\frac{t}{t_0} = \int_0^{R/R_0} dx~ x^{d/2} e^x, 
\label{eq:radius_soln}
\ee
where $t_0^{-1} = \sqrt{E_0 [N (1-r^2)]^{d+2}/(\rho_0 \Omega_d 
2^{d+1})}$ and $R_0^{-1}= N (1-r^2)/2$.

For later reference, it will be useful to derive the asymptotic 
solutions to Eq.~(\ref{eq:radius_soln}).  Let $\alpha= \ln(t/t_0)$. Then 
for large times, it is straightforward to derive:
\be
\frac{R}{R_0}=
\alpha \left[1-\frac{d}{2} \frac{\ln\alpha}{\alpha} + \frac{d}{2} 
\frac{\ln\alpha}{\alpha^2}
+O \left(\frac{1}{\alpha^2}\right)
\right], ~\alpha \gg 1.
\label{eq:radius_largetime}
\ee
The growth is logarithmic at large times in all dimensions. For short 
times, by writing the exponential in Eq.~(\ref{eq:radius_soln}) as a 
series, it is easy to obtain
\be
\frac{R}{R_0}= \left[\frac{(d+2) t}{2 t_0}
\right]^{\frac{2}{d+2}} \left[ 1 + O \!
\left( \left( \!\! \frac{t}{t_0}\right)^{\frac{2}{d+2}} \right) \right] \!, t
\ll t_0.
\label{eq:radius_shorttime}
\ee
For small times, the power law growth of radius is identical to the 
elastic case \cite{taylor}.

The experimental data in BCK was fitted to the numerical solution of 
Eq.~(\ref{eq:radius_soln}) with $d=2$. Although the equation describes 
the data well (see Fig.~4 of Ref.~\cite{boudet2009}), we now argue that 
the analysis has certain shortcomings, making the results questionable.

First, we show by a simple calculation that the solutions 
Eqs.~(\ref{eq:energy_soln}) and (\ref{eq:radius_soln}) do not give the 
correct results when $d=1$. The solution Eqs.~(\ref{eq:energy_soln}) and (\ref{eq:radius_soln}) are valid for all values of $r<1$, including $r=0$. In one dimension, the special case $r=0$, when particles stick on collision, is easily solvable \cite{zahera}. Let particles 
of mass $m$ be initially placed equidistant from each other with 
inter-particle spacing $a$. Pick a particle at random and give it a 
velocity $v_0$ to the right. When this particle collides with its 
neighbor, it coalesces with it. After $k$ collisions, the mass of the 
composite particle is $(k+1) m$, its distance from the impulse is $R=k 
a$, and its velocity, given by momentum conservation, is $v_k=v_0/(k+1)$ 
towards the right. The time taken for $k$ collisions is given by
\be
t_k= \sum_{i=0}^{k-1} \frac{a}{v_i}=
\frac{a k(k+1)}{2 v_0}.
\ee
Solving for $k$, we obtain $k=(-1+\sqrt{1+ 8 t v_0/a})/2$. At large 
times $t \gg a/v_0$, the radius and energy are $R = ka \approx \sqrt{2 
v_0 a t}$ and $E(t) = m v_0^2 a/(2 R)$.  The analysis in BCK for energy 
[Eq.~(\ref{eq:energy_soln})] and radius R(t) 
[Eq.~(\ref{eq:radius_largetime})] are not consistent with the exact 
solution in one dimension.

Second, there is a conserved quantity in this system, even though energy 
is not conserved. Every collision is momentum conserving. In addition, 
the clustering of all the displaced particles at the rim of the ring 
prevents momentum being transfered in the negative radial direction. If 
we further assume that once the dense rim is formed, the angular 
coordinates 
of particles do not change much, then radial momentum is a constant of 
motion (see also discussion on Fig.~\ref{radmom} in
Sec.~\ref{bckmodel_numerics}).  Therefore,
\be
\Omega_d R(t)^d V(t) \Delta\theta = \rm{constant}, 
\label{eq:radialmomentum}
\ee
where $\Delta\theta$ is the angular spread in direction $\theta$. The 
solution to Eq.~(\ref{eq:radialmomentum}) is
\be
R(t) \propto t^{\alpha}, ~t \gg t',
\label{eq:radius_scaling}
\ee
where $\alpha=1/(d+1)$, and $t'$ is the initial mean collision time. 
Equivalently $R(t)^{d/2} \sqrt{E(t)}$ is a constant of motion. 
Equation~(\ref{eq:energy_soln}) is clearly not consistent with this 
constraint, neither is Eq.~(\ref{eq:radius_largetime}) for growth of 
radius consistent with Eq.~(\ref{eq:radius_scaling}).

We, therefore, conclude that the analysis of BCK is not completely 
satisfactory. Since the solution of BCK [Eqs.~(\ref{eq:energy_soln}) and 
(\ref{eq:radius_soln})] was based on the assumption that $N(t)$, the 
rate of collisions per particle per unit distance, is a constant, we 
test the validity of this assumption as well as the prediction of 
Eq.~(\ref{eq:radius_soln}) in molecular dynamics simulations of a hard 
sphere gas. As we will argue below, the theory presented in BCK is 
also applicable to the hard sphere model.

\section{\label{model} Model for Simulation}

The system which is simulated is defined as follows.  Consider a 
collection of identical particles, modeled as hard spheres, in two dimensions. 
The mass and diameter of the particles are set to unity. All the 
particles are initially at rest and have a packing density $0.20$, much 
smaller than the known random closed packed density $0.84$ in two dimensions
\cite{williams,bideau1984}.  We model an isotropic impulse by 
introducing four particles at the center with speed $v_0$ in the 
directions $0, \pi/2, \pi$, and $3\pi/2$. Particles interact only on 
collision, during which momentum is conserved and velocities change 
deterministically. If the velocities of two particles $1$ and $2$ 
before and after collision are 
${\bf u}_1$, ${\bf u}_2$, and ${\bf v}_1$, ${\bf v}_2$ respectively, 
then
\be 
{\bf v}_{1,2}={\bf u}_{1,2}-\epsilon [{\bf n}.({\bf u}_{1,2} -{\bf u}_{2,1})] 
{\bf n}, 
\ee 
where $r=2 \epsilon-1$, $ (0<r<1)$ is the coefficient of restitution and 
${\bf n}$ is the unit vector directed from center of particle $1$ to the 
center of particle $2$.  In words, the tangential component of the 
relative velocity remains unchanged, while the magnitude of the 
longitudinal component is reduced by a factor $r$. When $r=1$, the collisions are elastic. To simulate the 
inelastic system, the coefficient of restitution $r$ is chosen to be 
less than unity if the magnitude of the longitudinal component of the 
relative velocity is greater than a velocity scale $\delta$, else $r=1$ 
mimicking elastic collisions for small relative velocities 
\cite{bennaim1999}. This qualitatively captures the experimental 
situation where $r$ is seen to be a function of the relative velocity 
\cite{raman,fauve}. In addition, it prevents inelastic collapse that is 
a hindrance to simulations \cite{Mcnamara,McNamara1994}. The cutoff 
$\delta$ introduces a timescale in the problem at large times, after 
which most of the collisions are elastic. For sufficiently small 
$\delta$, the elastic crossover timescale does not show up in our 
simulations.

We simulate the system in two dimensions using event driven molecular 
dynamics \cite{rapaportbook}. The data presented are typically averaged 
over 8 different initial realizations of the particle configurations.  All lengths are 
measured in units of the particle diameter, and time in units of initial 
mean collision time $t_0=v_0^{-1} n^{-1/d}$, where $v_0$ is the initial 
speed and $n$ is the number density. The value of $\delta$ is $10^{-4}$, 
unless specified otherwise. For these values of $\delta$, all the 
quantities that we measure except for the rate of collisions are 
independent of $\delta$ \cite{zahera}. The initial speed is $v_0=1$ 
unless specified otherwise.

\section{\label{bckmodel_numerics} Numerical study of BCK results}
\begin{figure}
\begin{tabular}{cc}
\includegraphics[width=0.48\columnwidth]{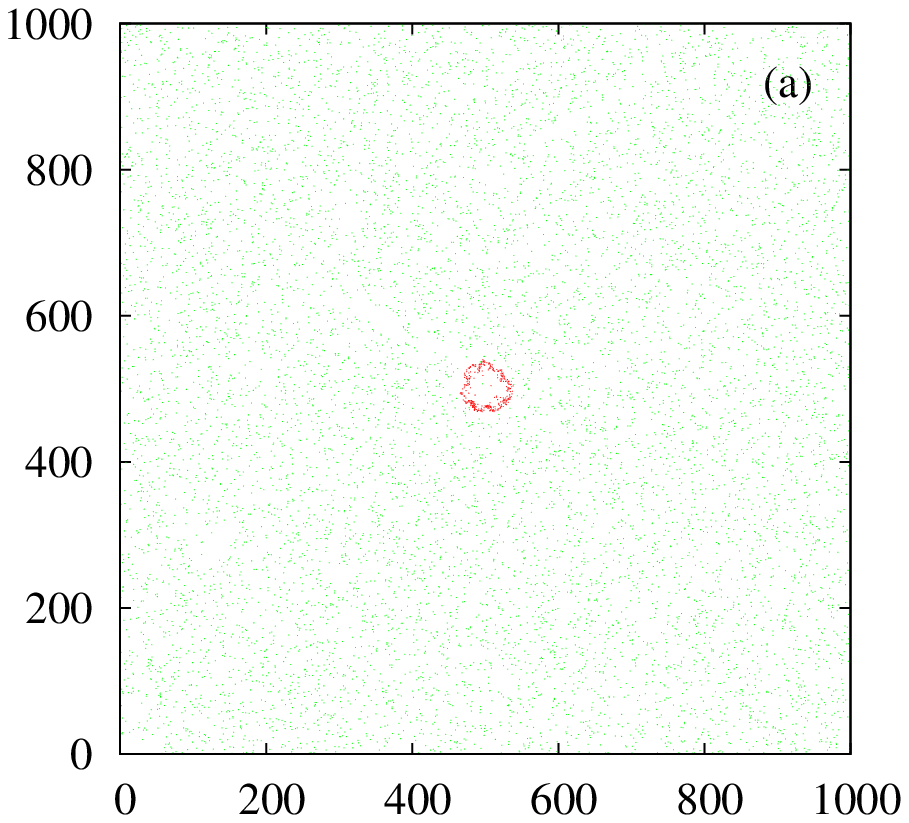}
\includegraphics[width=0.48\columnwidth]{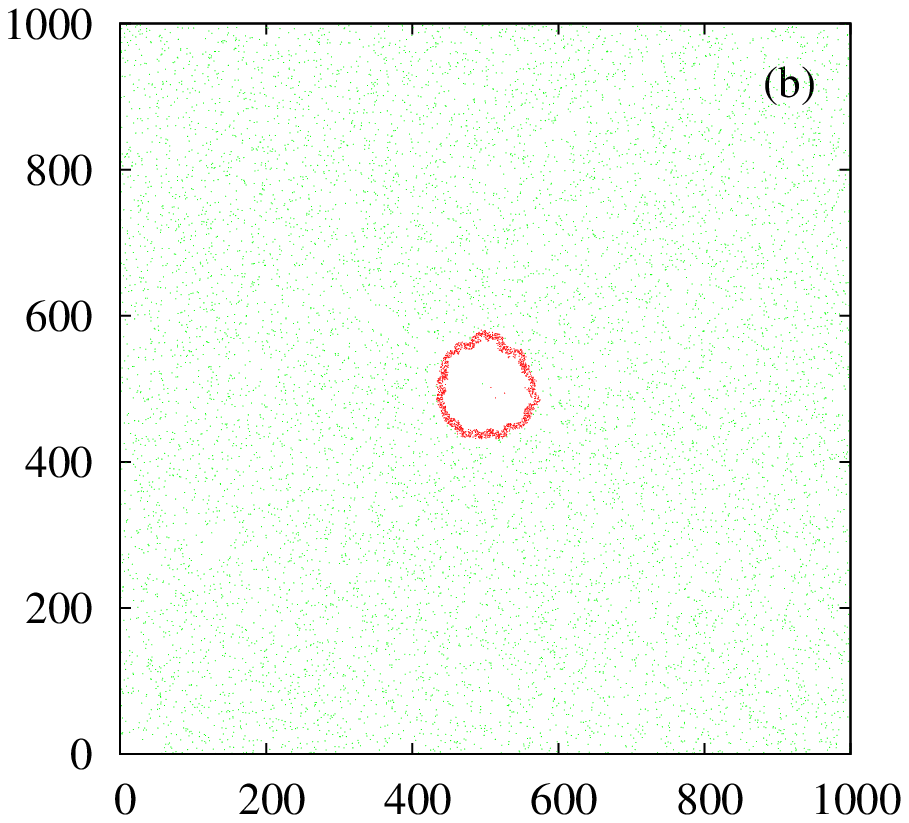}\\
\includegraphics[width=0.48\columnwidth]{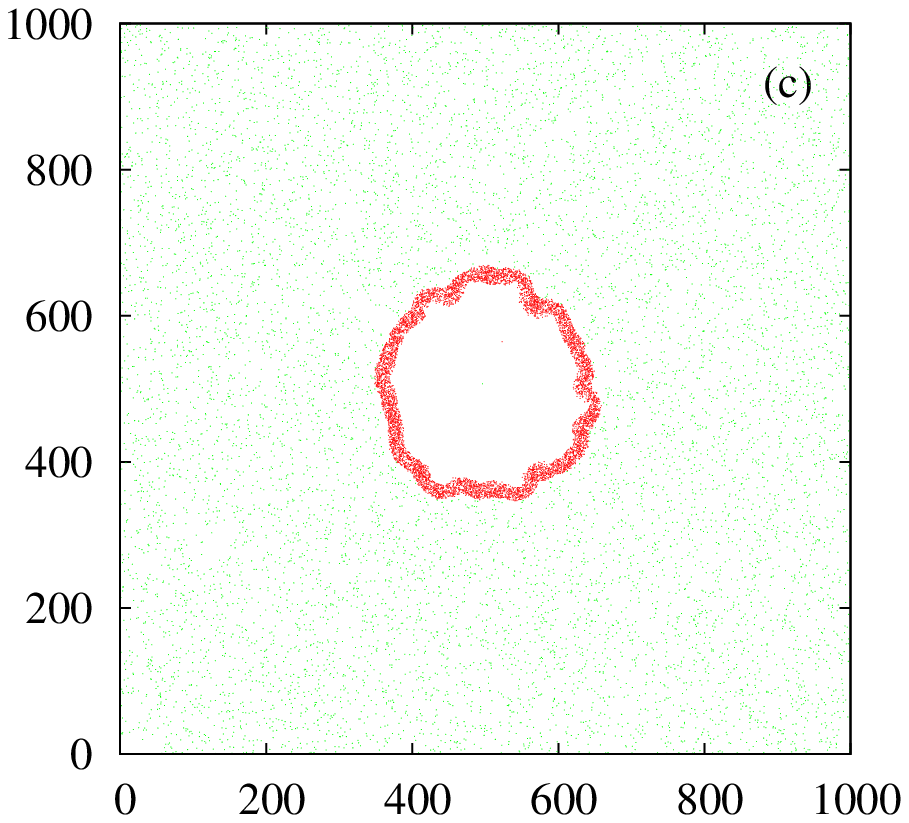}
\includegraphics[width=0.48\columnwidth]{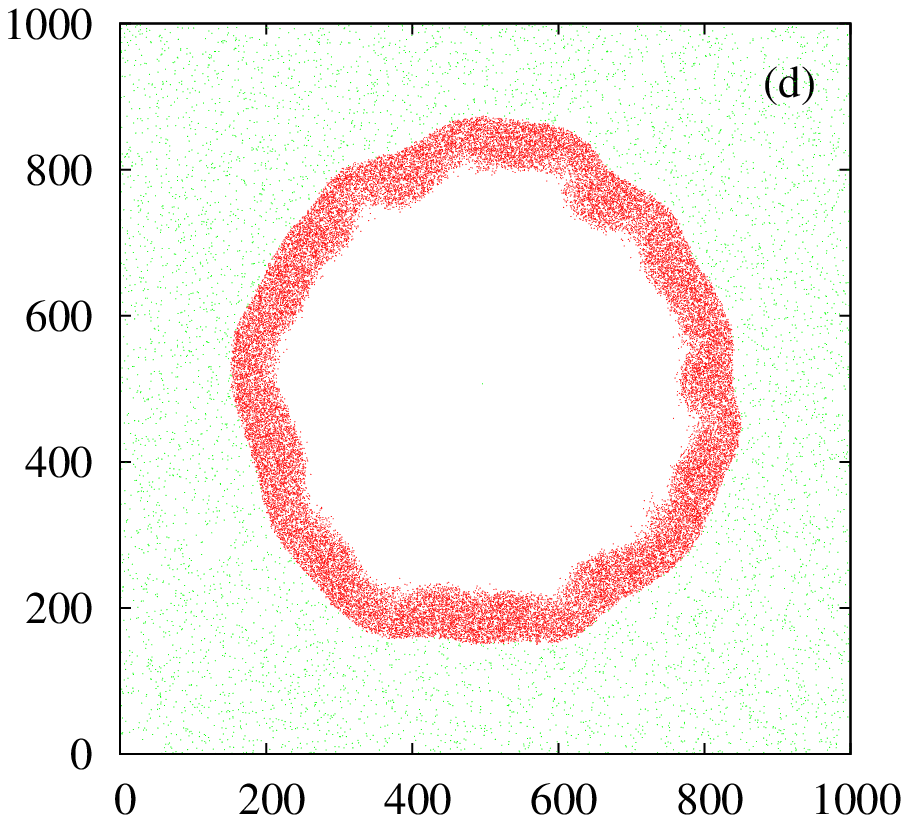}\\
\end{tabular}
\caption{(Color online) Moving (red) and stationary (green) particles at times 
$t=$ (a) $10^{3}$, (b) $10^{4}$, (c) $10^{5}$ and  (d) $10^{6}$,
following an isotropic impulse at $(500, 500)$ 
at $t=0$. The moving particles cluster together at the disturbance front. The data are for $r=0.10$.}
\label{fig1}
\end{figure}

In this section, the results and assumption of BCK are checked in numerical simulation of the hard sphere model. In Fig.~\ref{fig1}, we show the time evolution of the system following 
an impulse. As time increases, all the particles that were originally in 
a roughly circular ring, cluster together at its rim. We observe 
clustering for all the values of $r<1$ that we have simulated, with 
clustering setting in at later times for larger coefficients of 
restitution.

The formation of an empty region bounded by the moving particles (as in 
Fig.~\ref{fig1}) is the only requirement for the BCK theory to be 
applicable. Therefore, if the analysis in BCK is correct, then the results for radius in 
Eq.~(\ref{eq:radius_soln}) should describe the disturbance in the hard 
sphere model too. In numerical simulations, data can be obtained for 
much longer times than that in the experimental data in 
Ref.~\cite{boudet2009}, and therefore be used to make a more rigorous 
test of the assumptions and the conclusions of the BCK theory.

\begin{figure}
\includegraphics[width=\columnwidth]{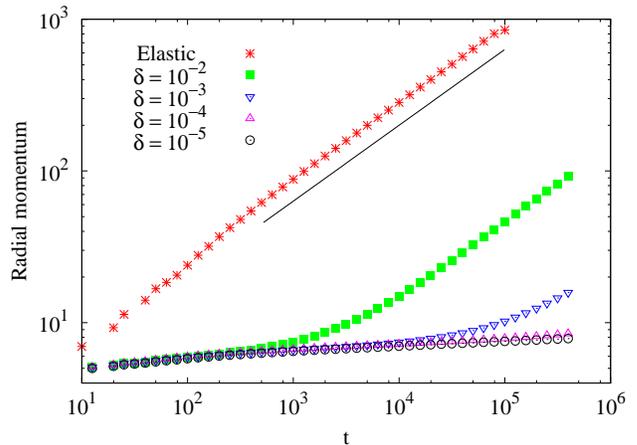}
\caption{\label{radmom}(Color online) The radial momentum as a function of time. For elastic collisions, it
increases as $\sqrt{t}$ (the solid straight line is a power law 
$\sqrt{t}$). For inelastic collisions with $r=0.10$, the radial momentum is a constant at large times when $\delta \rightarrow 0$.
}
\end{figure}

We first present data confirming that radial momentum is a constant of 
motion, as argued in Sec.~\ref{bckmodel}. In Fig.~\ref{radmom}, 
the temporal variation of the radial 
momentum is shown for different $\delta$ with fixed $r=0.10$ and 
compared with the data for the elastic problem. When all collisions are 
elastic, radial momentum increases as $\sqrt{t}$. When collisions are inelastic, radial 
momentum tends to a constant at large times as $\delta \rightarrow 0$.

\begin{figure}
\includegraphics[width=\columnwidth]{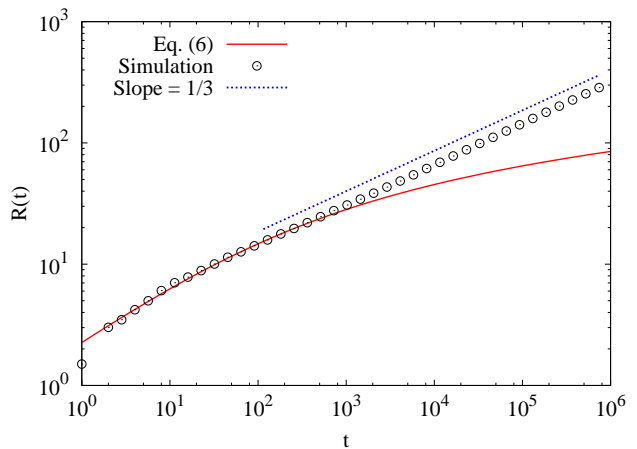}
\caption{\label{fig_radius}(Color online) Data for radius  $R(t)$ from simulations in two dimensions 
are  compared with Eq.~(\ref{eq:radius_soln}) and $t^{1/3}$. $R_{0}$ and $t_{0}$ in Eq.~(\ref{eq:radius_soln}) are obtained by fitting the initial time simulation data to Eq.~(\ref{eq:radius_soln}) and are $R_{0} = 10.30 \pm 0.21$ and $t_{0} = 35.79 \pm 2.35$.
The data are for $r=0.10$.
}
\end{figure}

In Fig.~\ref{fig_radius}, we compare the BCK result 
Eq.~(\ref{eq:radius_soln}) for the radius with hard sphere simulation 
data. The constants $R_0$ and $t_0$ in Eq.~(\ref{eq:radius_soln}) are 
determined by fitting it to the numerical data at early times. It is 
clear that Eq.~(\ref{eq:radius_soln}) captures only the short time 
behaviour. On the other hand, the data at large times are consistent with 
the power law $t^{1/3}$. We believe that the discrepancies between the 
short and large time behaviour are not brought out by the experimental 
data as the time scales are not large enough.

\begin{figure}
\includegraphics[width=\columnwidth]{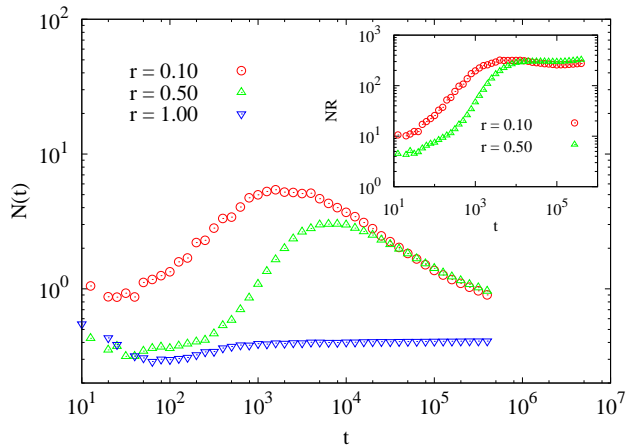}
\caption{\label{fig_collision} (Color online) Temporal variation of $N(t)$, the number of 
collisions per particle per unit distance for various $r$. 
For $r<1$, $N(t)$ is not constant as assumed by BCK. Inset:
$NR$, where $R$ is the radius of disturbance, is a constant at large times for $r<1$. }
\end{figure}

We now make a direct test of the BCK assumption that $N(t)$, the number 
of collisions per particle per unit distance is a constant in time, as 
assumed in BCK. The data for $N(t)$ are shown in Fig.~\ref{fig_collision} 
for three different coefficients of restitution, one of them being 
$r=1$. While $N(t)$ is a constant when collisions are elastic, it is 
clearly not so for $r<1$, invalidating the BCK assumption. At large 
times, the rate of collisions become independent of $r$ as long as 
$r<1$. This is consistent with the observations in the freely 
cooling granular gas \cite{bennaim1999,nie2002}, where the long time 
behavior of $E(t)$ and $N(t)$ is independent of $r$, and hence identical 
to $r=0$, the sticky limit. Thus we could think of the rim as a solid 
annulus made up of all the particles that have undergone at least one 
collision. Therefore, once the rim 
forms, we expect that only the collisions of the particles that 
are at the outer edge of the rim, with the stationary particles are 
relevant. Then, the collisions per particle on surface per unit time, $N R$,
should be constant. This is confirmed in the inset of 
Fig.~\ref{fig_collision}, where $N R$ tends to a constant independent of 
$r$, at large times. Since the relevant collisions are taking place at the outer boundary 
of the rim, Eqs.~(\ref{eq:energy_soln}) and (\ref{eq:radius_soln}) 
underestimate the radius, or equivalently overestimate energy loss.

\section{\label{comparison} Comparison with experimental data}

\begin{figure}
\includegraphics[width=\columnwidth]{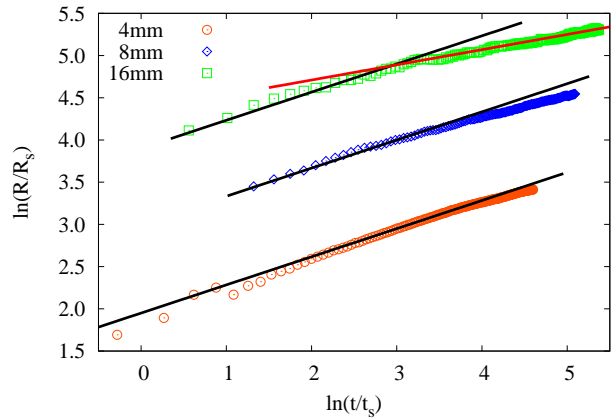}
\caption{\label{fig_comparison}(Color online) Experimental data from 
Ref.~\cite{boudet2009} for radius $R$ as a function of time $t$ 
following an impact by steel balls of diameter $4$ mm, $8$ mm and $16$ 
mm. The black/red lines have slope $1/3$ and $0.18$ 
respectively. $R_s$ is the diameter of a glass bead and $t_s$ is the 
mean time taken by a glass bead to traverse a distance equal to its 
diameter. The data have been obtained from Ref.~\cite{boudet2009}.
}
\end{figure}

In this section, we compare the power law solution $R(t) \sim t^{1/3}$, 
obtained from the conservation of radial momentum, with the experimental 
data of Ref.~\cite{boudet2009}.  Figure~\ref{fig_comparison} shows the 
data (Fig.~4 of Ref.~\cite{boudet2009}) for the temporal variation of 
the radius of disturbance $R(t)$ following impacts with spheres of 
different diameter. The black solid lines are power laws $t^{1/3}$. 
There are temporal regimes where it matches well with the experimental data. 
However, there are deviations from $t^{1/3}$ at large times. There is 
sufficient statistics for this late time regime only for the impact with 
the largest sphere. For this data, we find that the data are best fitted 
by a power law $t^{0.18}$ (see green line in Fig.~\ref{fig_comparison}).

The experimental situation is more complicated than the simple hard 
sphere model for which the power law growth is presumably the correct 
result. To equate the two, we had to make approximations. First, we 
ignored the fluctuations of the velocities of the particles about the 
mean velocity. While this is reasonable for large impact velocities when 
typical speeds of displaced particles are much larger than typical velocity 
fluctuations, the fluctuations become relevant at late time. Second, we 
ignored the experimentally observed three dimensional nature of the rim 
(see discussion in last but one paragraph of Ref.~\cite{boudet2009}). 
Such a possibility will result in radial momentum not being conserved, 
thus invalidating the scaling arguments in \cite{zahera}.

It is possible that either or both of these approximations could be 
responsible for the crossover seen at large times. In 
Sec.~\ref{modifications}, we study modified versions of the hard sphere 
model, which incorporates the above features. We argue that the 
crossover from $t^{1/3}$ law can be explained by these models.

\section{\label{modifications} Effect of non-zero ambient temperature
and three dimensional rim}

In the center of mass coordinates, all particles are not stationary but 
fluctuating about their mean position. When these velocity fluctuations 
become comparable to the velocity of the rim, then we expect the rim to 
destabilize, and power laws to show crossovers.

We model this situation as follows. Initially all the particles (type 
$E$) are assumed to be elastic and equilibrated at a certain fixed 
temperature, parametrized by $\Lambda^2 = \langle v^2 \rangle/ v_0^2$, 
where $\langle v^2 \rangle$ is the mean velocity fluctuations and $v_0$, 
as earlier, is the speed of the perturbed particles. $\Lambda=0$ 
corresponds to the case when all particles are initially stationary. An isotropic impulse is 
imparted by introducing four particles (type $I$) at the center with speed $v_0$ in the 
directions $0, \pi/2, \pi$, and $3\pi/2$. 
Collisions between $E$ particles are elastic. Collisions involving at 
least one $I$ particle are inelastic. If an $E$ particle collides with 
an $I$ particle, then it becomes type $I$. This model captures shock 
propagation in a system where all particles have some nonzero kinetic 
energy.

\begin{figure}
\begin{tabular}{cc}
\includegraphics[width=0.48\columnwidth]{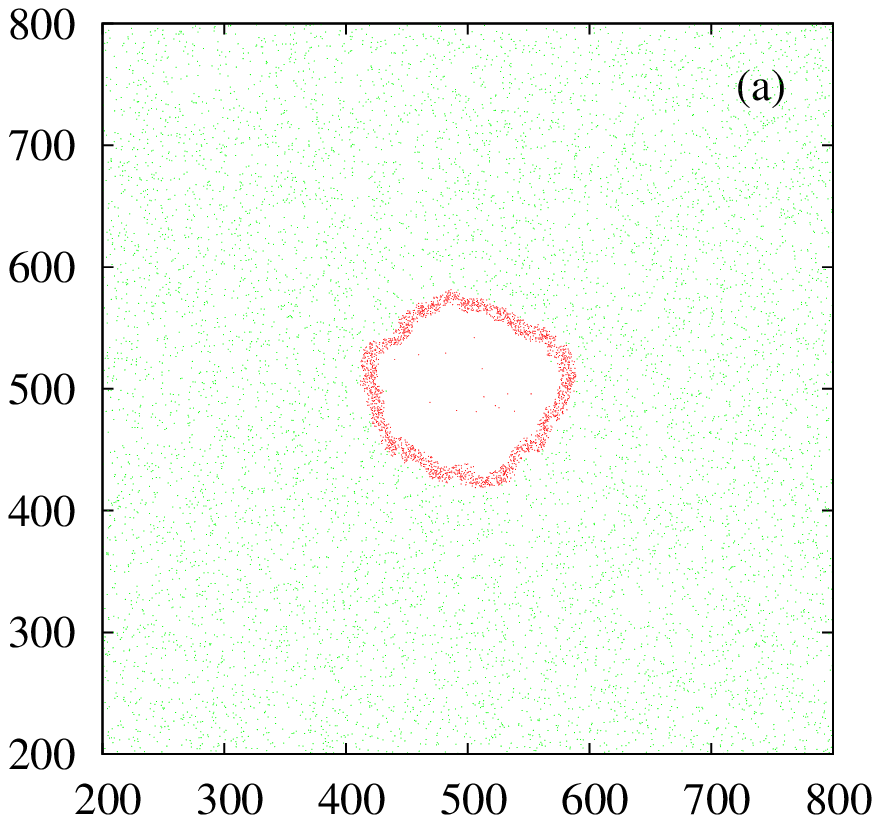}&
\includegraphics[width=0.48\columnwidth]{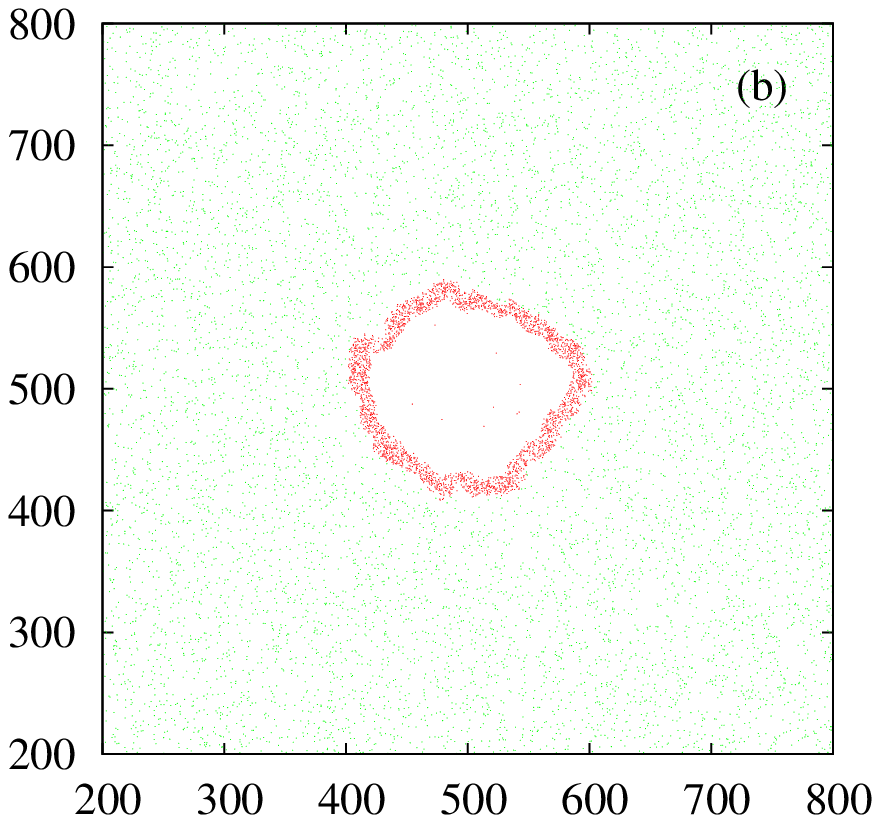}\\
\includegraphics[width=0.48\columnwidth]{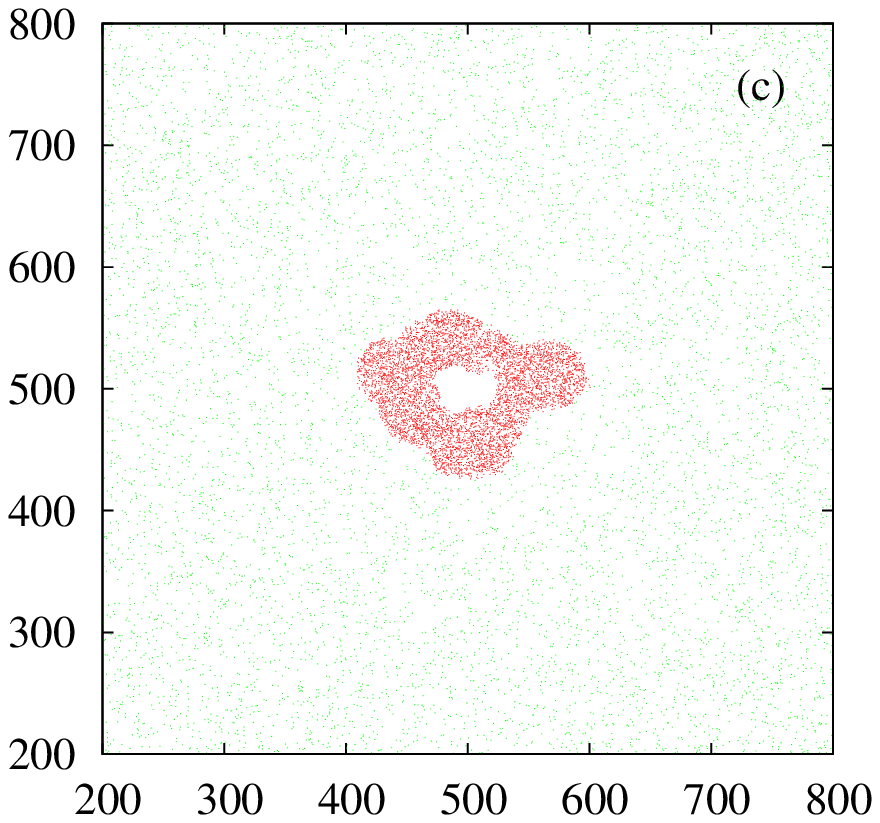}&
\includegraphics[width=0.48\columnwidth]{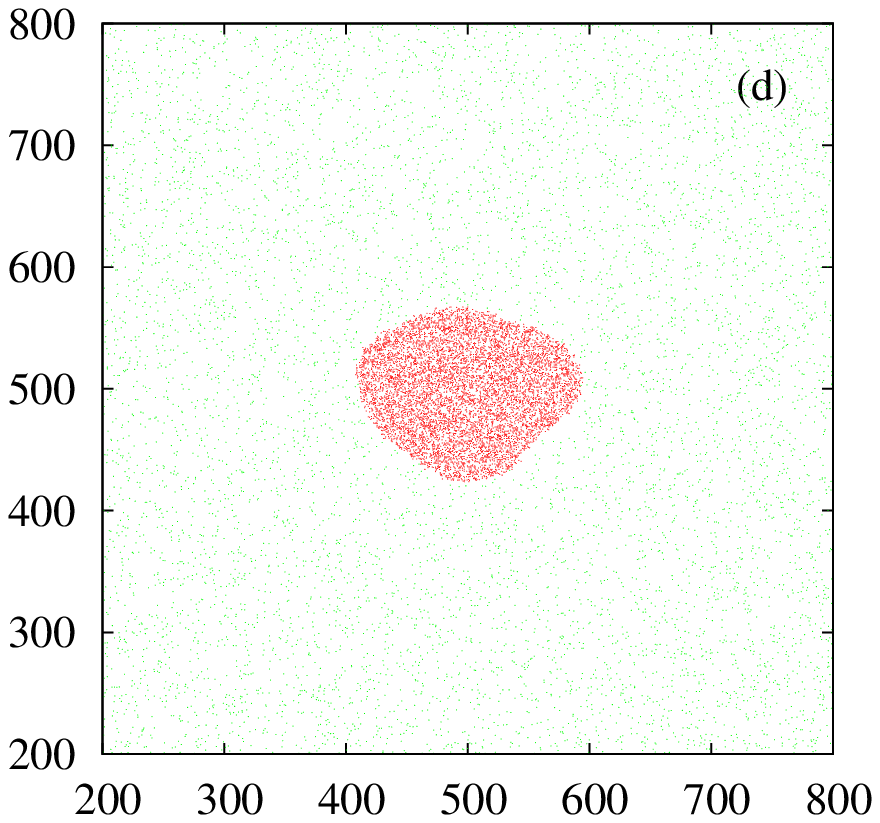}
\end{tabular}
\caption{\label{fig_snapshots} (Color online)
Snapshots of inelastic I particles (red) and elastic E particles (green),
when $\Lambda = 1/800$, following an isotropic impulse at $(500, 500)$ at 
$t=0$. The time increases from (a) to (d) and correspond to the times
shown by labels a--d in Fig.~\ref{amb_radius}.
Initially, the disturbance grows as in Fig.~\ref{fig1}, but at late times due to
velocity fluctuations, the rim gets destabilized.
The data are for $r=0.10$.}
\end{figure}

In Fig.~\ref{fig_snapshots}, we show snapshots of the system at various 
times, when the $\Lambda=1/800$. The sharp rim starts becoming more 
diffuse as the velocity of the rim decreases, until the enclosed empty 
region vanishes completely. These snapshots are qualitatively very 
similar to that seen in the experiment for low speed impacts and at large times (see Fig.~1 of 
Ref.~\cite{boudet2009}).

\begin{figure}
\includegraphics[width=\columnwidth]{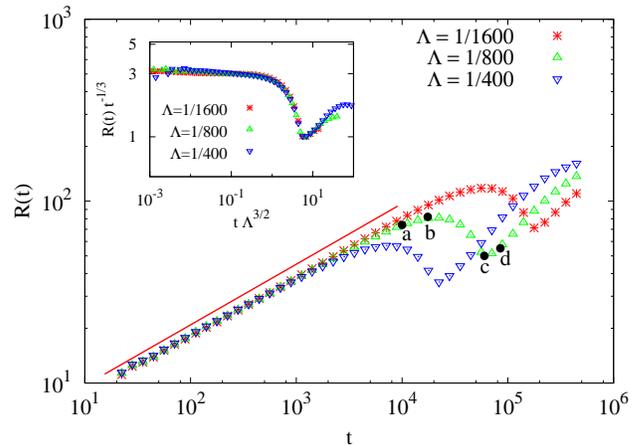}
\caption{\label{amb_radius}(Color online) The radius of disturbance $R(t)$ as a function of
time $t$ for different values of $\Lambda$. The effect of velocity
fluctuations are felt later for smaller $\Lambda$. Inset: Data
collapse when scaled according to Eq. (\ref{lambdascaling}). 
A solid red line of slope 1/3 is drawn for reference. The data are for $r=0.10$.
}
\end{figure}

When the rim destabilizes, $R(t)$ shows deviation from the $t^{1/3}$ 
power law growth (see Fig.~\ref{amb_radius}). It is straightforward to 
estimate this crossover time $t_c$. The instability sets in when the 
speed of the rim is of the same magnitude as the velocity fluctuations, 
i.e. $v_{t_c} \sim \Lambda v_0$. Since $v_{t} \sim dR/dt \sim t^{-2/3}$, we 
immediately obtain $t_c \sim \Lambda^{-3/2}$. Thus, $R(t)$ should have 
the scaling form
\be
R(t) \sim t^{1/3} f\left(t \Lambda^{3/2} \right),
\label{lambdascaling}
\ee
where $f(x)$ is a scaling function with $f(x) \sim O(1)$, when $x
\rightarrow 0$. The curves for different $\Lambda$ collapse when
scaled as in Eq.~(\ref{lambdascaling}) [see inset of
Fig.~\ref{amb_radius}].

The introduction of a finite ambient temperature, while leading to the 
disintegration of the rim, does not produce the large time behaviour of 
the data for the radius. We now ask whether the rim becoming three 
dimensional could be responsible for that. The rim presumably becomes 
three dimensional because a fast particle when hemmed in by many 
surrounding particles may jump out of the plane due to collision with 
floor and friction. The net effect is a reduction in radial momentum, 
which could change the growth law.
\begin{figure}
\includegraphics[width=\columnwidth]{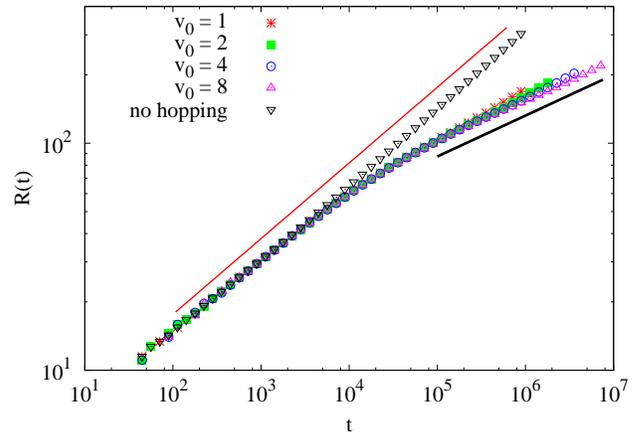}
\caption{\label{hopp_radius}(Color online) Temporal variation of radius $R(t)$ for $\kappa = 
0.20$ with various initial velocity $v_{0}$. The black line is a power law $t^{0.18}$ while the red line is a power law $t^{1/3}$. The data with no hopping correspond to $v_{0}=1$. All data are for $r=0.10$.}
\end{figure}

We consider the following model.  We divide the system into squares of 
length equal to diameter of the particles. Given the grid position of a 
particle, any particle which is in one of the eight neighboring squares 
will be called its neighbor. At any instant of time, if a particle has 
eight or more neighbors, then we remove the particle if its velocity 
${\bf v}$ satisfies the hopping criterion, $({\bf v} - {\bf v}_{cm}). 
{\bf \hat{v}}_{cm} > \kappa v_{cm}$, where ${\bf v}_{cm}$ is the center 
of mass velocity of the particle and its neighbors. In words, the 
longitudinal component of the velocity should be larger than $v_{cm}$ by 
a factor $\kappa$.

The hopping criterion is tested for all moving particles after every 
$100$ collisions in the system, and the results do not depend on this 
number provided it is not too large. The results are shown in 
Fig.~\ref{hopp_radius}. The results obtained are insensitive to the 
value of $\kappa$ provided $\kappa <0.20$. We find that at large times, 
the system crosses over to a different power law growth $\approx 
t^{0.18}$, that is very similar to the growth law seen in the 
experiment. While the aim of the model was to show that loss of 
radial momentum, at high densities, can result in crossovers at large 
times, we obtain a quantitative match.  As of now, we have no 
explanation why the exponents have approximately the same numerical 
value, and it could be just a coincidence.

\section{\label{conclusion} Conclusion and discussion}

In summary, we analyzed the recent experiment \cite{boudet2009} of 
dropping spheres onto a flowing monolayer of glass beads.  We modeled 
the experiment with a hard sphere system undergoing inelastic 
collisions. With this hard sphere system, we showed that the assumption 
of constant rate of collision per particle per unit distance, made in 
the theory \cite{boudet2009} to describe the experimental data is 
correct only for elastic particles. For inelastic system, the
relevant collisions are the collisions of the particles at the outer edge 
of the rim with the stationary particles outside. We also argued that
the formation of the circular ring in the perturbed system conserves 
radial momentum. This conservation law leads to a $t^{1/3}$ 
power law growth for the radius of disturbance. The $t^{1/3}$ growth law 
describes the experimental data well except at large times when the 
data show a crossover to a different power law growth.  We 
attributed this crossover to the rim becoming three dimensional because 
of high densities and collisions with the floor. By constructing a 
simple model incorporating these effects, we were able to explain the 
crossovers at large times.

The current experimental data can not distinguish between the theory in
BCK and the power law growth argued for in this paper. If the
experimental time scale is increased, then such a distinction may be 
possible.  It will be worthwhile to make the attempt.

In our simulations, we modeled the coefficient of restitution as $r<1$
for relative velocities larger than a velocity scale $\delta$ and
$r=1$ otherwise.  The velocity scale $\delta$ is relevant experimentally
and not just a computational tool. Experimentally,  $r(v)$ 
approaches $1$ when the relative velocity $v$
tends to zero, i.e., $1-r(v) =g(v/\delta)$, where
$g(x) \sim x^\chi+ O(x^{2 \chi})$, for $x \ll 1$ and $g(x) \sim
O(1)$ for $x\rightarrow \infty$.
Experimentally, the exponent $\chi$ takes a variety of values.
Within the framework of
viscoelastic theory, $\chi=1/5$ \cite{brilliantovbook}. 
Systems with $\chi<1$ cannot be
studied using the event driven molecular dynamics simulations performed in
this paper, as inelastic collapse
prevents the simulation from proceeding forward.
However, we have checked, using molecular dynamics simulations with soft
potentials, that the rim formation and radius increasing as a power law
$t^{1/3}$ continue to be true for $\chi <1$ \cite{sudhir_dae}. 

It will be quite interesting to see if any connection can be made 
between the shock problem in which most of the particles are initially
stationary and the well studied freely cooling granular 
gas, in which all particles initially have a nonzero kinetic energy. 
It may be possible to think of the freely cooling gas as a collection
of shocks initiated at different points in space, which interact when
the shock fronts meet. If such a connection is possible, it will help in resolving the
uncertainty of the energy decay 
exponent \cite{nie2002,carnevale} of the freely cooling granular gas.
Thus, it will be useful to make a detailed study 
of the case of two interacting shocks.

The data for radius show a crossover from an initial elastic behavior
$t^{1/2}$ to an asymptotic  $t^{1/3}$ growth law. It would be of interest
to understand this crossover better. Exact solution of the shock problem
in one dimension  with $0 < r < 1$ would throw light on it. An exact solution appears possible given that the freely cooling in one dimension is one of the exactly solvable model in granular physics. 

\begin{acknowledgments}
We thank the authors of Ref.~\cite{boudet2009} for providing us with
the experimental data. These simulations were carried out at
ANNAPURNA at the Institute of Mathematical Sciences.
\end{acknowledgments}


\end{document}